\documentclass[aps,prl,groupedaddress,twocolumn]{revtex4}

\usepackage{amsmath,amsfonts,amssymb,graphics,graphicx,epsfig,color,times,bbm,natbib}

\begin{document}

\title{Quantum Communication with Quantum Dot Spins}

\author{Christoph Simon$^1$} \email{christoph.simon@ujf-grenoble.fr}
\author{Yann-Michel Niquet$^2$}
\author{Xavier Caillet$^1$}
\author{Jo\"{e}l Eymery$^2$}
\author{Jean-Philippe Poizat$^1$}
\author{Jean-Michel G\'{e}rard$^2$}
\affiliation{$^1$ Laboratoire de
Spectrom\'{e}trie Physique, CNRS
- Universit\'{e} de Grenoble 1, Saint Martin d'H\`{e}res, France\\
$^2$ Service de Physique de Mat\'{e}riaux et Microstructures,
DRFMC, CEA, Grenoble, France}

\date{\today}

\begin{abstract}
Single electron spins in quantum dots are attractive for
quantum communication because of their expected long
coherence times. We propose a method to create entanglement
between two remote spins based on the coincident detection
of two photons emitted by the dots. Local nodes of two or
more qubits can be realized using the dipole-dipole
interaction between trions in neighboring dots and spectral
addressing, allowing the realization of a quantum repeater.
We have performed a detailed feasibility study of our
proposal based on tight-binding calculations of quantum dot
properties.
\end{abstract}


\maketitle

Implementations of quantum information protocols in the
solid state are of interest because they may eventually be
more scalable than other approaches. Individual electron
spins in quantum dots \cite{lossdivince} are a promising
system for quantum computing and quantum communication due
to their expected long coherence times. Spin relaxation
times as long as 20 ms have been observed at 4 T and much
longer times predicted for lower magnetic fields
\cite{kroutvar}. There are theoretical predictions that in
the absence of nuclear spins the decoherence time of the
spins might approach their relaxation time \cite{golovach}.
Nuclear spins can be eliminated completely e.g. by using
isotopically purified II-VI materials, since Zn, Cd, Mg, Se
and Te all have dominant isotopes without nuclear spins.

For quantum commmunication it is important to be able to
create entanglement between remote spins
\cite{leuenberger,childress}. The recent proposal of Ref.
\cite{leuenberger} relies on achieving a large Faraday
rotation for a single photon due to the quantum dot spin.
It requires very high-finesse micro-cavities that are
switchable in a picosecond. The proposal of Ref.
\cite{childress} relies on the detection of a single photon
that could have been emitted by either of two remote
sources \cite{cabrillo}. This approach is attractive
because it does not require a finely controlled strong
spin-photon interaction. A practical drawback of the scheme
of Ref. \cite{childress} is the requirement of phase
stability over the whole distance. Ref. \cite{simonirvine}
proposed a scheme that creates entanglement between two
remote emitters via the detection of two photons, which
eliminates this stability requirement, while keeping the
advantages of an emission-based scheme. In the present work
we demonstrate, firstly, how to realize a similar scheme
for quantum dot spins. Secondly, we show that it is
possible to realize local nodes of two or more spins using
dipole-dipole interactions and spectral addressing. Such
nodes allow the realization of quantum repeater protocols
\cite{repeater,childress}. We have investigated the
feasibility of our proposal in detail, including numerical
calculations of the electronic properties of quantum dots
using tight binding methods.

Our scheme applies to flat quantum dots, such as typical
strain-induced quantum dots or dots in heterostructured
nanowires \cite{nanowires}. This implies that the
lowest-energy hole states will have predominantly
``heavy-hole'' character, and will be well separated from
predominantly ``light-hole'' states. The dots can be
charged with single electrons via tunneling controlled by
an electric field as in Ref. \cite{karraicharging}. A
magnetic field is applied in growth direction. The qubit
states are the two spin states corresponding to the lowest
electron level in the dot, denoted by $|1/2\rangle$ and
$|-1/2\rangle$. We use transitions between the qubit states
and the two lowest-energy trion states, which have angular
momentum 3/2 and -3/2, see Fig. 1. A trion consists of the
electron that is trapped in the dot plus an exciton (i.e.
an electron-hole pair created by the incoming light). The
two electrons form a spin singlet, the angular momentum of
the trion is therefore determined by that of the hole,
which is $\pm 3/2$. Note that a Lambda system like in Ref.
\cite{simonirvine} could be realized by applying a
transverse magnetic field. However, we prefer the
configuration with the field in growth direction because it
makes it much easier to realize qubit measurements and
two-qubit gates.

\begin{figure}
\begin{center}
\includegraphics[width=0.9 \columnwidth]{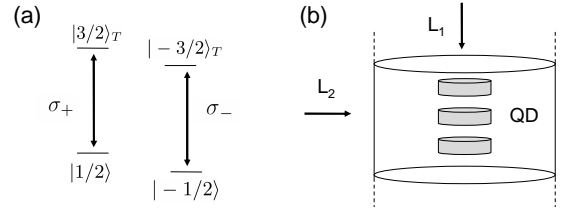}
\caption{(a) Level scheme underlying entanglement creation, qubit
measurement and two-qubit gates. The $|1/2\rangle$ electron state
is coupled to the $|3/2\rangle_T$ trion state by $\sigma_+$
radiation propagating along the growth direction, while the
$|-1/2\rangle$ electron state is coupled to the $|-3/2\rangle_T$
trion by $\sigma_-$ radiation. The other transitions, which have
$|\Delta J|=2$, are strongly suppressed, cf. below. The photon
energies are $E_{\sigma_{\pm}}=E_T \pm E_Z$, where $E_T$ is the
trion energy in zero field and $E_Z=g_X \mu_B B$ is the Zeeman
energy, with $g_X$ the $g$ factor of the exciton. (b) Outline of a
single node. A stack of quantum dots (QD) is embedded in a
waveguide. L$_1$ is a control beam addressing the transitions in
(a), a perpendicular beam L$_2$ is required for one-qubit gates.
The waveguide ensures efficient collection of the emitted
photons.} \label{structure}
\end{center}
\end{figure}

The protocol for entanglement creation starts by creating a
superposition state
$\frac{1}{\sqrt{2}}(|1/2\rangle+|-1/2\rangle)$ of the spin
via a single-qubit rotation as described below. Then one
applies simultaneous $\pi$ pulses to both the $|1/2\rangle
\rightarrow |3/2\rangle_T$ and the $|-1/2\rangle
\rightarrow |-3/2\rangle_T$ transitions, creating the state
$\frac{1}{\sqrt{2}}(|3/2\rangle_T+|-3/2\rangle_T)$. This
state will decay under photon emission, creating an
entangled spin-photon state $\frac{1}{\sqrt{2}}(|1/2\rangle
|\sigma_+\rangle+|-1/2\rangle|\sigma_-\rangle)$, where the
photon states $|\sigma_+\rangle$ and $|\sigma_-\rangle$
differ not only in polarization but also in energy.  Such
spin-photon entangled states are created for two remote
quantum dots $A$ and $B$, which have been carefully tuned
such that $E_{\sigma_+}^A=E_{\sigma_+}^B$ and
$E_{\sigma_-}^A=E_{\sigma_-}^B$. In this case the photons
from $A$ and $B$ will be indistinguishable for each
polarization, which makes it possible to perform a partial
Bell state analysis on them using only linear optical
elements \cite{bsa,simonirvine}. The method is based on the
fact that only the antisymmetric state $|\sigma_+\rangle_A
|\sigma_-\rangle_B-|\sigma_-\rangle_A |\sigma_+\rangle_B$
leads to coincidences between the two output ports if both
photons are combined on a beam splitter. The required
two-photon interference occurs even if the photon energies
corresponding to the two polarizations are different (in
contrast to what is implied in Ref. \cite{simonirvine}).
The emitted photons can be collected efficiently and guided
to the location of their joint measurement using waveguides
and optical fibers; e.g. Ref. \cite{zhang} obtained a
coupling coefficient of 95 \% for a monochromatic emitter
inside a single-mode waveguide. The Bell measurement of the
photons projects the two remote spins into an entangled
state \cite{simonirvine}.

It is important that the photon emission is coherent. This
is possible for resonant excitation as described. The
experiment of Ref. \cite{langbein} showed exciton dephasing
times longer than 30 ns in InAs quantum dots. For a
realistic radiative lifetime of 300 ps this would imply
dephasing related errors below the 1 \% level. Resonant
excitation requires separating the pump light from the
photons that one wants to detect. This can be done
temporally using a fast electro-optic switch. There are
currently available Pockels cells with switching times
shorter than 100 ps, which would already be enough to
detect most of the desired photons.

A deviation from the conditions
$E_{\sigma_+}^A=E_{\sigma_+}^B$ and
$E_{\sigma_-}^A=E_{\sigma_-}^B$ by an amount $\delta E$
will lead to an error in the Bell measurement due to
imperfect wavefunction overlap of $(\delta E)^2/\gamma^2$,
where $\gamma$ is the inverse of the radiative lifetime.
For a lifetime of 300 ps, one needs a precision of 0.2
$\mu$eV for an error of 1\%. To achieve both conditions,
one has to be able to tune both the trion energy in zero
field $E_T$, which can be done by varying the temperature
\cite{varshni}, and the Zeeman energy $E_Z$, which can be
done by varying the magnetic field. The required precision
of control can be estimated to be of order 5 mK for the
temperature and of order 1 mT for the magnetic field. These
values are realistic with present technology.

For a combined collection and detection efficiency
$\eta=0.25$ for each photon, the proposed scheme allows to
entangle two spins separated by 20 km in 8 ms, which is the
same time as is obtained for the scheme of Ref.
\cite{childress} with the same $\eta$ and an emission
probability of 8 \%. Note that the latter probability has
to be kept relatively small for the protocol of Ref.
\cite{childress} to avoid errors due to the emission of two
photons.

The superposition
$\frac{1}{\sqrt{2}}(|1/2\rangle+|-1/2\rangle)$ required for
entanglement creation can be realized via Raman transitions
exploiting the fact that there are excited trion states
that have significant dipole moments with both electronic
ground states \cite{calarco,nazir}. For our chosen field
configuration, one of the two laser beams has to propagate
in a direction orthogonal to the growth axis. A detailed
scheme for realizing arbitrary one-qubit operations via
Raman transitions is described in Ref. \cite{chen}. The
most important error mechanism is the decoherence of the
excited trion state. The related error can be estimated
\cite{caillet} to be below $10^{-3}$ for a decoherence rate
$\gamma=3 \times 10^{10}/$s as in Ref. \cite{htoon} and a
realistic detuning of order 30 meV. Coherent manipulation
of single spins in quantum dots via Raman transitions has
recently been demonstrated experimentally \cite{dutt}.

Qubit measurements can be realized via cycling fluorescence
as proposed in Ref. \cite{shabaev}. If $\sigma_+$ radiation
is applied in resonance with the $|1/2\rangle \rightarrow
|3/2\rangle_T$ transition, and the electron is originally
in the $|1/2\rangle$ state, then the system will cycle
between the $|1/2\rangle$ and $|3/2\rangle_T$ states
emitting photons, whereas no photons will be emitted if the
electron is originally in state $|-1/2\rangle$. The
occurrence of ``forbidden'' transitions from
$|3/2\rangle_T$ to $|-1/2\rangle$ limits the number of
cycles that can be used for detection. However, the
forbidden transitions are strongly suppressed in quantum
dots with high cylindrical symmetry. In our numerical
calculations on cylindrical quantum dots in a nanowire
structure, cf. below, we found probabilities for the
forbidden transition at the level of $10^{-3}$ per cycle,
which allows of order $10^3$ cycles. For experimental
results showing precise optical selection rules in
self-assembled quantum dots see Ref. \cite{cortez}. In
practice, a mean number of 20 detected photons in
combination with a threshold of 10 counts for a positive
detection of the bright state will ensure that measurement
errors are below 0.5 \%.

We will now describe how to implement local nodes
consisting of two or more interacting spins. We propose to
realize local two-qubit gates based on spin-selective
excitation combined with the dipole-dipole interaction
between trions in neighboring quantum dots, using a
fixed-detuning variation of the protocol of Ref.
\cite{calarco}. One again applies $\sigma_+$ radiation
close to resonance with the transition from $|1/2\rangle$
to $|3/2\rangle_T$. An excitation will thus only happen if
the electron is in state $|1/2\rangle$. If a trion is
excited in the neighboring dot as well, an additional phase
is accumulated due to the dipole-dipole interaction. The
two spins acquire this phase only if they are both in the
state $|1/2\rangle$, which makes it possible to realize a
controlled phase gate. A phase due to the dipole-dipole
interaction between excitons in a pair of quantum dots has
recently been observed \cite{unold}. To enhance the
interaction, the trions can be made to have permanent
dipoles by applying an electric field orthogonal to the
growth direction. For example, for two stacked flat quantum
dots whose centers are separated by 10 nm, an electron-hole
separation of 5 nm gives a dipole-dipole interaction energy
$E_{dd}$ of order 5 meV. There are different techniques for
fabricating stacked quantum dots that are sufficiently
close together. One promising approach is the use of
heterostructured semiconductor nanowires as in Refs.
\cite{nanowires}.

The gate operation is performed adiabatically, i.e. the
exciting laser is slightly detuned from the trion
resonance.  An important source of error for the two-qubit
gates is spontaneous emission of photons from the trion
state, the probability of which is $\Gamma \int dt P_T(t)$,
where $\Gamma$ is the spontaneous decay rate and $P_T(t)$
is the population in the trion state. For example, choosing
a laser Rabi frequency $\Omega(t)=\Omega_0 e^{-t^2/\tau^2}$
with $\Omega_0=1.0 \times 10^{12}/$s, $\tau=11$ ps and a
detuning $\Delta=0.75 \times 10^{12}/$s gives a controlled
phase of $\pi$. For these values $\int dt P_T(t)$ is equal
to 3.4 ps, which would give a 1.1 \% error for a trion
lifetime of 300 ps as considered above. This error is
reduced to 0.34 \% for a lifetime of 1 ns, for which the
coherence and control requirements discussed above still
appear realistic.

\begin{figure}
\begin{center}
\includegraphics[width=0.6 \columnwidth]{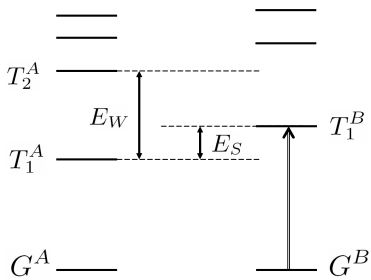}
\caption{Requirements imposed by spectral addressing. $G$
is the quantum dot ground state, $T_1$, $T_2$ etc. are the
trion states. Zeeman sublevels are not shown, i.e. $G$
corresponds to the states $|\pm 1/2\rangle$ and $T_1$ to
$|\pm 3/2\rangle_T$. Suppose that $A$ is the dot with the
lowest energy for $T_1$, and $B$ another dot in the same
node. When exciting the $G^B \rightarrow T_1^B$ transition,
one has to avoid exciting $G^A \rightarrow T_2^A$. This
defines an energy window $E_W$ in which $T_1^B$ has to lie.
On the other hand, $T_1^B$ has to be larger than $T_1^A$ by
at least $E_S$ in order to avoid exciting $T_1^A$ while
emitting a phonon.} \label{structure}
\end{center}
\end{figure}

For the dipole-dipole interaction to be effective, the dots
have to be very close together. Individual dots have to be
addressed spectrally. This is possible exploiting the fact
that the trion energies vary with the dimensions of the
dot. The number of qubits per node is restricted by the
existence of excited trion states, which imposes an energy
window $E_W$ for the trion energies $E_{\sigma_{\pm}}^K$
that can be used for addressing the qubits in a given node,
and by the interaction with phonons, which requires a
minimum energetic separation $E_S$ between dots, cf. Fig.
2.

A typical quantum dot has several excited electron and hole
states. While the electronic states are typically quite
well described by the effective mass approximation for the
electron (particle in a box), this is not the case for the
hole states. In order to be able to make quantitative
estimates, we have performed numerical calculations of the
electronic properties for quantum dot structures in a model
system. We have studied quantum dots of various dimensions
that are constituted by layers of GaAs in an
Al$_{0.4}$Ga$_{0.6}$As nanowire with circular cross
section. In order to simplify the discussion, we will focus
on a particular example, namely a dot with 16 nm diameter
and 4 nm thickness that is completely embedded in AlGaAs.
Such a structure can be fabricated by performing the radial
overgrowth \cite{lieber} of an AlGaAs shell layer over a
GaAs/AlGaAs axial nanowire heterostructure
\cite{nanowires}. The one-particle states of the nanowires
were computed in a tight-binding framework \cite{niquet},
using the $sp^3d^5s^*$ model of Ref \cite{jancu}. The
lowest-lying electron and hole wavefunctions of the $\simeq
750000$ atoms supercell were computed with a
Jacobi-Davidson algorithm as described in Ref.
\cite{niquet}. A transverse electric field of 5 mV/nm is
applied to introduce an electron-hole separation of 5 nm.
The first four excited hole states for the described dot
lie 15, 24, 26 and 30 meV above the hole ground state (not
counting Zeeman sublevels). The first excited electron
state is 48 meV above the electron ground state. The first
excited trion state therefore consists of the electron in
its ground state and the hole in its first excited state.
In the presence of the electric field, the strength of this
transition is about 1/4 of the lowest energy trion
transitions, i.e. it is far from negligible. This implies
that the energies of the (lowest-lying) trions for all dots
in a node should lie in a window $E_W$ of order 15 meV, cf.
Fig. 2.

Light that is in resonance with $T_1^B$ can excite $T_1^A$
while emitting an acoustic phonon, cf. Fig. 2. Following
Refs. \cite{takagahara,krummheuer} one can show that the
rate for this process is given by $\gamma(\Delta,t)=2 \pi
J(\Delta)\Omega^2(t)/\Delta^2$, where $\Delta$ is the
detuning, $\Omega(t)$ is the Rabi frequency of the laser
and the function $J(\Delta)=\frac{\Delta^3}{16 \pi^3 \rho
c^5} \int d^2{\bf n} |D(\Delta {\bf n}/c)|^2$. Here $\rho$
is the density, $c$ the sound velocity, the integral is
over the surface of the unit sphere, and $D({\bf k})=\int
d{\bf r} [D_v |\psi_v({\bf r})|^2-D_c |\psi_c({\bf
r})|^2]\exp(-i {\bf k}\cdot {\bf r})$, where $D_c$ and
$D_v$ are the deformation potential constants for the dot
material \cite{takagahara}. The wave functions $\psi_v$ and
$\psi_c$ are those of the hole and electron ground states
respectively. The wave functions are obtained by the
tight-binding calculations described above. For the above
choice of $\Omega(t)$, one finds that a separation of
$E_S$=7.5 meV (corresponding to the center of the energy
window, since $E_W$=15 meV for our example) reduces the
error due to phonon emission to 0.14 \%. Together with the
error due to spontaneous emission of 0.34 \% predicted
above, this means that it is possible to realize nodes
containing two qubits such that the total error for
two-qubit gates is of order 0.5 \%. Three qubits per node
are possible if one tolerates a total error of order 2 \%.
We have focused on the two-qubit example in order to
facilitate comparison with Ref. \cite{childress}, which
shows that a quantum repeater protocol with two-qubit nodes
and local errors for two-qubit gates and measurements (cf.
above) of 0.5 \% is rather efficient. For example, it would
allow to establish an entangled pair over 1000 km in a few
seconds, with neighboring nodes separated by 20 km, as in
our above discussion. We have thus shown that our proposed
scheme is capable of the same performance, without the
requirement of phase stability for the optical fiber links.

We studied the model system of GaAs in AlGaAs because all
the relevant parameters are sufficiently well known to make
quantitative predictions. However, equivalent results are
to be expected for appropriate II-VI systems \cite{adachi},
which have the advantage of allowing the elimination of
nuclear spins, as mentioned above. For example, for ZnSe
the effective hole masses are about 50 \% larger than for
GaAs, which might lead to a proportionately smaller energy
window. However, the deformation potential constants are
predicted to be significantly smaller than for GaAs, which
would lead to a smaller required separation for the same
dot dimensions. The chosen dot dimensions are the result of
an (informal) optimization. Reducing the dot dimensions,
e.g., increases the level separation (and thus $E_W$), but
it also makes the function $J(\Delta)$ wider, and thus
increases $E_S$.

The spectral addressing requirements for the qubit
measurements are less severe than for the two-qubit gates
because the necessary light intensities are smaller. For
the one-qubit gates, since the Raman lasers are far detuned
from the trion energies, the trion resonances cannot be
used to address individual dots. However, since for the
Raman process the difference in laser frequencies has to be
equal to the energy difference between the two qubit
states, one can use the variation in Zeeman energies
between individual dots for addressing. The electron $g$
factors vary with the size of the quantum dots. Published
results on self-assembled dots in III-V and II-VI systems
\cite{nakaoka,prado,pryorflatte} suggest that it is quite
feasible to achieve a variation in Zeeman energy of order 1
$\mu$eV in a 1 T magnetic field for dots whose trion
energies differ by 7.5 meV. This is consistent with gate
times for the one-qubit gates below 10 ns, with negligible
addressing errors.

We have shown how to create entanglement between remote
quantum dot spins by first entangling the spins with
photons emitted by the dots, and then detecting the two
photons in the Bell basis. We have demonstrated that it is
possible to realize local nodes consisting of two or more
quantum dots such that nearest neighbors are coupled by
dipole-dipole interactions between excitons. Based on a
detailed study of expected errors and a comparison with the
results of Ref. \cite{childress}, we have shown that our
proposed protocol should allow the realization of efficient
quantum repeaters without requiring interferometric
stability.

This work was supported by the EU Integrated Project {\it Qubit
Applications (QAP)}. We would like to thank L. Besombes, A.
Bychkov, S. Hastings-Simon, F. Jelezko, O. Krebs, and T. Stace for
helpful comments and discussions.

\bibliographystyle{apsrev}

\end{document}